\documentclass[nonatbib]{ragtime} 

\usepackage{epsf}

\title[Optical spectroscopy of Cyg X-1]%
      {Optical spectroscopy of Cyg X-1}

\author[P.~Hadrava]
     {Petr Hadrava 
\\
    \ins{}Astronomical~Institute, Academy~of~Sciences, Bo\v{c}n\'{\i}~II~1401,\splitins[1]
        CZ-14131~Prague, Czech~Republic\\
}

\begin{document}


\begin{abstract}
The star HDE 226868 known as an optical counterpart of the black hole
candidate Cyg X-1 has been observed in H$_\alpha$ region using
spectrograph at Ond\v{r}ejov 2-m telescope. The orbital parameters
are determined from He\,I-line by means of the author's method of Fourier
disentangling. Preliminary results are also presented of disentangling
the H$_\alpha$-line into a P-Cyg profile of the (optical) primary
and an emission profile of the circumstellar matter (and a telluric
component).
\end{abstract}

\section{Introduction}
The bright X-ray source Cyg X-1 has been identified with the
star denoted as HDE 226868, V1357 Cyg or BD+34$^{\circ}$3815 etc.
An improvement of instrumentation of the Ond\v{r}ejov 2-m telescope
enabled to start with systematic observations of this target of
magnitude V$\simeq$8.9, B$\simeq$9.6. With coordinates
$\alpha_{2000}=19^{h}58^{m}21.7^{s}$, $\delta_{2000}=+35^{\circ}12'6''$
it is well observable from Ond\v{r}ejov mainly at summer.

 It is known to be an interacting binary with period $P\simeq 5.6 d$.
The primary component is a supergiant of spectral type classified as
B0 (or O9.7) Iab  with temperature $T_{\rm eff}=30400\pm500$ K and
log $g=3.31\pm0.07$. This primary, which nearly fills its Roche lobe,
shows signs of variable strong stellar wind and an overabundace of He
and heavier elements (cf. e.g. Karitskaya et al. 2007).

 The secondary component invisible in optical radiation is a compact
object, most probably a black hole. This companion, or its neighborhood
emits a variable X-radiation, which is supposed to originate
from an accretion disk fed by the stellar wind from the primary.

 The X-radiation switches chaotically between two states. In the low/hard
state the total X-ray flux is low and the spectrum is flat, so that
the hard tail of X-radiation prevails. In the high/soft state the soft
radiation is enhanced more, and consequently the spectrum has a steeper
decrease toward the higher energies and hence the radiation is softer
in the mean. Some intermediate states may also appear temporarily.

 The X-ray flux is anticorrelated with the strength of emission in
the H$_{\alpha}$-line of the primary: in the X-low/hard state the
H$_{\alpha}$ emission is strong, while in the X-high/soft state
the H$_{\alpha}$ emission is weak.

 The aim of the observational campaign at Ond\v{r}ejov observatory was
to improve orbital parameters of the system, to check a possible
spectroscopic features connected with the circumstellar matter (either
accretion disk around the black hole, gaseous streams or stellar wind)
or with a possible third body, and to get line-profiles enabling a
quantitative comparison with a model of the atmosphere and stellar wind
of the primary. The first part of obtained spectra was provided for a
study on Cyg X-1 organized in a wide international collaboration,
the results of which should appear in Gies et al. (2007).
In the present contribution, results obtained using the author's
method of spectra disentangling from the same set of Ond\v{r}ejov
spectra are given. A more detailed study taking into account also
recently obtained spectra is in progress.

\section{Observational data}
The set of spectra used here consists of 24 exposures obtained with
CCD in the focus of 700mm camera of Coud\'{e} spectrograph of
the Ond\v{r}ejov 2-m telescope between April 1st and September 21st
2003. A typical resolution is about 0.25\AA{} per pixel.
The rough data have been processed by M. \v{S}lechta according to
\v{S}koda \& \v{S}lechta (2002).

\setlength{\unitlength}{1mm}
\begin{figure}[h]
\begin{picture}(127,63)
 \put(4.,-5){\epsfxsize=120mm\epsfbox{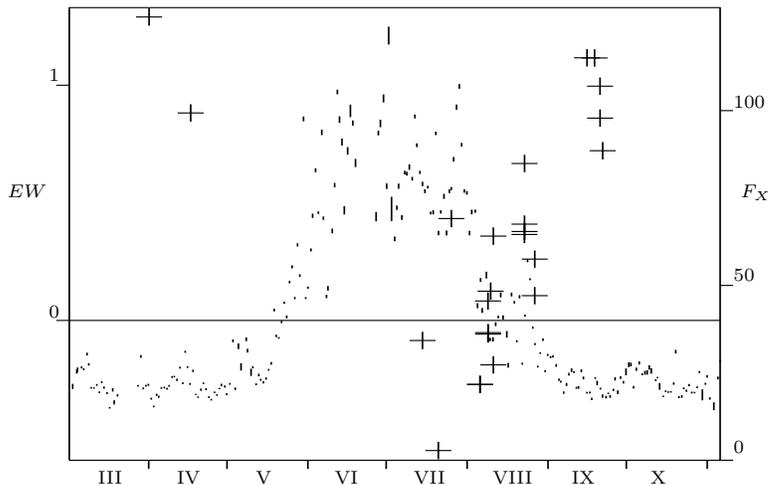}}
\put(24.5,0.){\scriptsize III}
\put(35.,0.){\scriptsize IV}
\put(45.5,0.){\scriptsize V}
\put(56.,0.){\scriptsize VI}
\put(66.5,0.){\scriptsize VII}
\put(77.,0.){\scriptsize VIII}
\put(87.5,0.){\scriptsize IX}
\put(98.,0.){\scriptsize X}
\put(109.,4.){\scriptsize 0}
\put(109.,27.){\scriptsize 50}
\put(109.,50.){\scriptsize 100}
\put(110.,38.){\scriptsize $F_{X}$}
\put(12.5,38.){\scriptsize $EW$}
\put(18.,22.){\scriptsize 0}
\put(18.,53.5){\scriptsize 1}
\end{picture}
\caption{
 \label{obr01} The equivalent width of H$_{\alpha}$ emission (in \AA{},
 crosses) in 2003 and ASM/RXTE one-day averages of sum-band intensity
 (in counts/s, error-bars)}
\end{figure}

 The observational period covers a transition of Cyg X-1 from the low
to high state and back, as it can be seen from the RXTE X-ray light-curve
in Fig.~\ref{obr01}.

 Examples of obtained spectra in both states are given in Fig.~\ref{obr02}.
It is obvious here that the upper spectrum taken at April before the
high-state episode has a strong emission in the whole H$_\alpha$ line
profile, while in August the emission remains in the long-wavelength
wing of the line only and the short-wavelength side of the line-profiles
reveals an absorption, as it is typical in the P-Cyg line-profiles of
stars losing mass via a stellar wind. The strength of the emission can
be quantified by the equivalent width of the line, i.e. by an integral
across the line of the intensity rectified to the continuum. These values
are plotted for each exposure in Fig.~\ref{obr01} which confirms the above
mentioned anticorrelation with the X-ray flux.

\setlength{\unitlength}{1mm}
\begin{figure}[hbt]
\noindent\begin{picture}(127,85)
 \put(3,-35){\epsfxsize=120mm\epsfbox{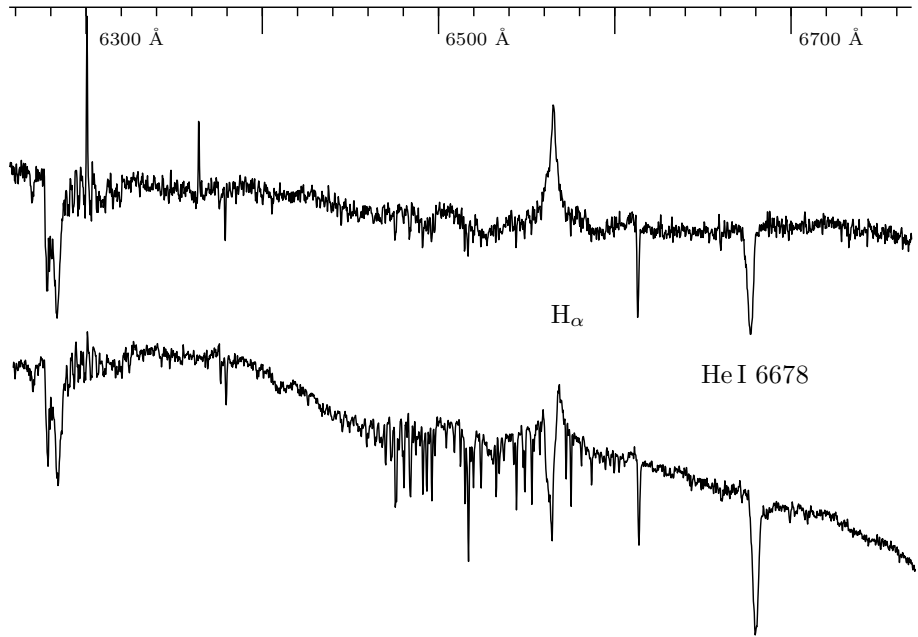}}
\put(75.,43.){H$_{\alpha}$}
\put(95.,35.){He\,I 6678}
\put(15.,80.){\scriptsize 6300 \AA}
\put(61.,80.){\scriptsize 6500 \AA}
\put(108.,80.){\scriptsize 6700 \AA}
\end{picture}
\caption{
 \label{obr02}Spectra of Cyg X-1 taken with Ond\v{r}ejov 2-m
 telescope on April 1st (upper curve) and August 5th (lower curve)
 2003 }
\end{figure}

 The He\,I-line 6678\AA{} is practically free of emission in both states.
It means that this line may enable to measure reliably radial
velocities of the primary component to get a constraint on the orbital
parameters of the system.

\section{Spectra disentangling}
Despite the radial velocities of the He\,I-line 6678\AA{} could be
measured using some standard method, it is advantageous to use the
author's method of Fourier spectra disentangling (cf. Hadrava 1994,
1997, 2004), which makes the procedure efficient and provides directly
the orbital parameters. The principle of the method (in the version of
1997 used here) consists in least-squares fitting of all the spectra
$I$ observed at various times $t$ as a superposition of unknown
spectra $I_{j}$ of the components in the form
\begin{equation}\label{Kor1}
 I(x,t;p)= \sum_{j=1}^{n}I_{j}(x)\ast s_{j}(t)\delta(x-v_{j}(t;p))\; .
\end{equation}
Here $x=c\,{\rm ln}\lambda$ is a logarithmic wavelength, $v_{j}$ are
instantaneous radial velocities of each component (or logarithms of redshift
$g$-factors for a general relativistic case), $s_{j}$ are factors fitting
possibly variable strengths of lines, $p$ are the orbital parameters to
be found. Fourier transform
\begin{equation}\label{FKor2}
 \tilde{I}(y,t;p)=
  \sum_{j=1}^{n}\tilde{I}_{j}(y)\;\tilde{\Delta}_{j}(y,t,p)
\end{equation}
of Eq.~(\ref{Kor1}) separates the solving for $I_{j}$ into individual
modes; similarly for $s_{j}$ one gets a set of linear equations, while
$p$ can be found by some numerical method of optimization (e.g. simplex
method in the author's code KOREL). Here $\Delta_{j}=s_{j}\delta(x-v_{j})$
in the present calculation, but generally it could also characterize some
general broadening function.

\setlength{\unitlength}{1mm}
\begin{figure}[hbt]
\noindent\begin{picture}(127,35)
 \put(0,0){\epsfxsize=70mm\epsfbox{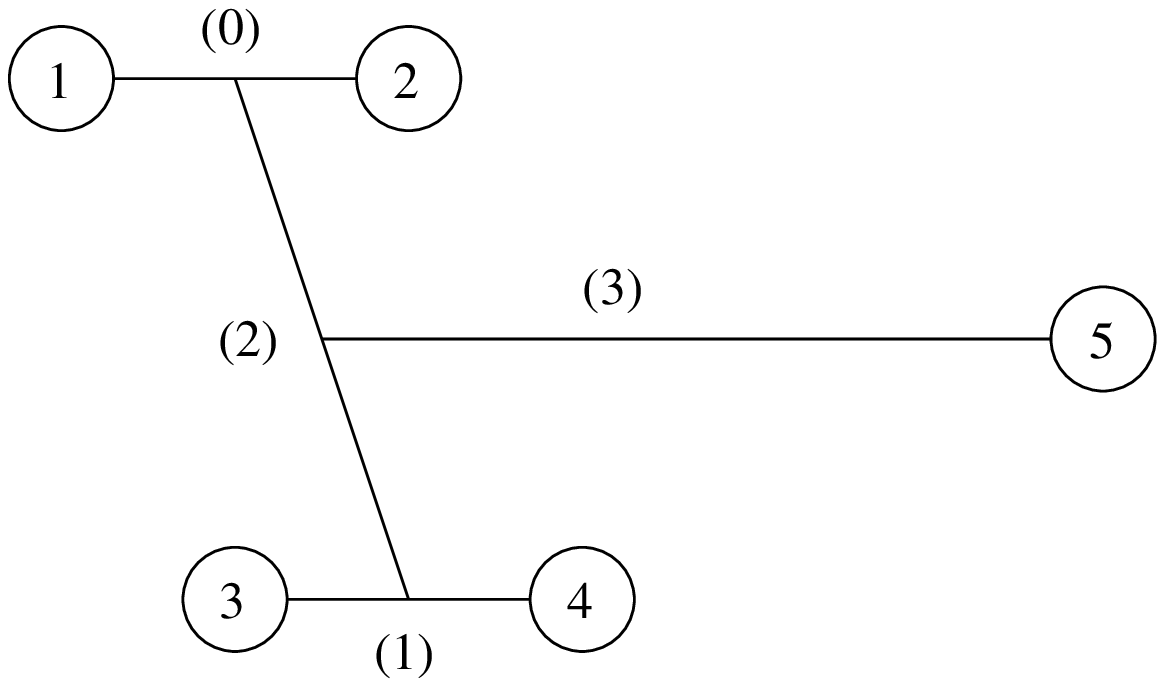}}
 \put(60,0){\epsfxsize=70mm\epsfbox{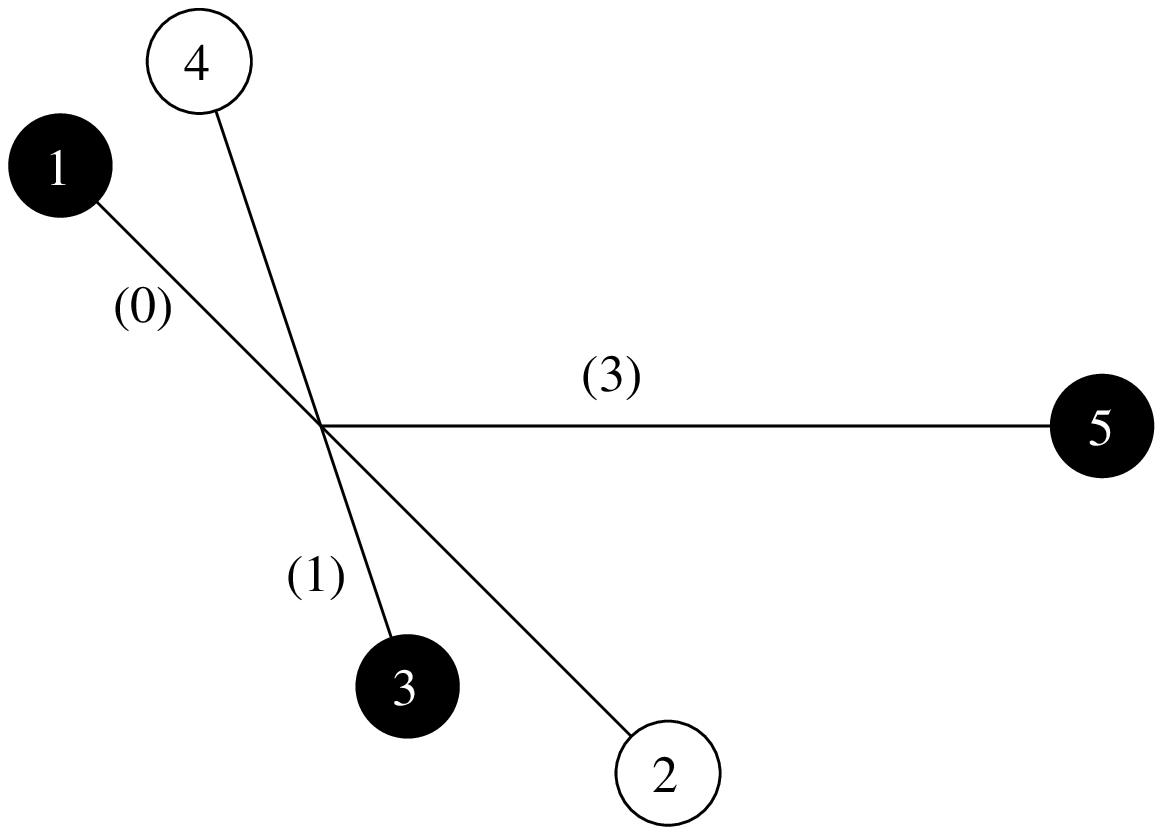}}
\end{picture}
\caption{
 \label{obr05} Hierarchical structure of a multiple stellar system
 supposed in KOREL-code (left) and its use for H$_{\alpha}$-line
of Cyg~X-1 (right)}
\end{figure}

 The method can generally provide intrinsic spectra of $n$ sources,
if more than $n$ observed spectra taken at different values $v_{j}$
are on input. Karas \& Kraus (1996) suggested a possibility to
disentangle in this way contributions of several spots to a line-profile
of an accretion disk. The author's code KOREL should be modified for
such a purpose, because it is designed for applications to systems
of binary or multiple stars. A hierarchical structure of the system is
supposed (cf. Fig.~\ref{obr05}, left), in which two pairs of close binaries
(denoted 1 + 2 and 3 + 4) are orbiting around their common centre of mass,
which may be on an even wider orbit with respect to another component
(Nr. 5). To be able to treat simpler systems, spectrum of each component
can be switched on or off, and the higher orbits (denoted by numbers in
parenthesis in Fig.~\ref{obr05}) may be chosen degenerated. At the same
time, this model is general enough to enable solving some more complicated
cases, e.g. just like a presence of circumstellar matter in a binary.

 As already mentioned, there are seen no traces of the companion or of
the circumstellar matter in the He\,I-line for Cyg X-1. Consequently, one
could take the extremally simple case $n=1$ of the disentangling (with
only the component 1 and orbit (0) switched on -- cf. Fig.~\ref{obr05},
right) for the spectral region around 6678\AA{}. However, because some
weak telluric lines are also present in this region, $n=2$ was used instead,
with the component 5 corresponding to the telluric lines and orbit (3) to
the annual motion. Results of the disentangling are shown on Fig.~\ref{obr03}.
In the upper part of this standard output from the KOREL-code, we can see
superimposed the 24 observed line-profiles (rectified to the continuum)
and their reconstructions from the disentangled line-profiles, which are
plotted as the two bottom curves. The disentangled values of orbital
parameters (the epoch and the amplitude of the radial-velocity curve)
are given in Tab.~\ref{tab1}.

\setlength{\unitlength}{1mm}
\begin{figure}[hbt]
\noindent\begin{picture}(127,55)
 \put(26,-20){\epsfxsize=75mm\epsfbox{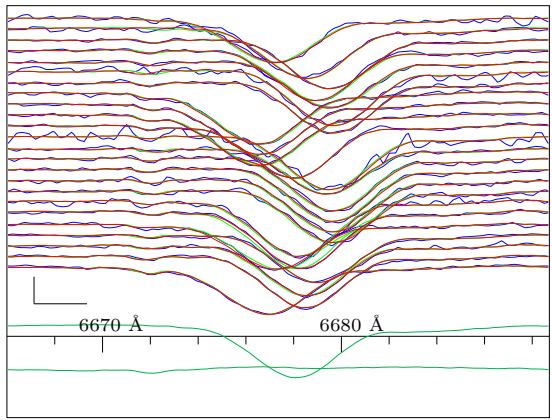}}
 \put(37.,10.3){\scriptsize 6670 \AA}
 \put(69.,10.3){\scriptsize 6680 \AA}
\end{picture}
\caption{
 \label{obr03} Disentangled line He\,I 6678 \AA}
\end{figure}

\section{Doppler mapping and disentangling of circumstellar matter}
Unlike the He\,I-line, the H$_{\alpha}$ shows the above mentioned
irregular emission, revealing a presence of strongly variable
circumstellar matter in the system. The variability of line-profile
of the emission component, which is frequent in many emission-line
systems, obviously violates the assumption on constancy of component
spectra $I_{j}$ in Eq.~(\ref{Kor1}) and makes the use of disentangling
for such systems questionable. On the other hand, one can always try,
if a violation of underlying assumptions is not an effect of second
order, and if a mean behaviour of the system cannot be approximated
neglecting this effect, or if the effect cannot be modelled as some
additional perturbation.

\begin{table}
\noindent{}\begin{tabular}{lccc}
 Line    & He\,I & H$_{\alpha}$  & H$_{\alpha}$ -- wind\\ \hline
 Period   & \multicolumn{3}{c}{ 5.599829 d }\\
 Periastron epoch & 52872.83 &  52873.01 & 52875.41\\
 Eccentricity  & \multicolumn{3}{c}{ 0.0}\\
 Periastron long. & \multicolumn{3}{c}{ --90$^{\circ}$}\\
 K$_{1}$ [km/s] & 71.94 &  71.26  &  60.78\\
\end{tabular}
\caption{\label{tab1} Disentangled orbital parameters of Cyg X-1}
\end{table}

Several attempts have been done by the author to fit discrepancies
between observed spectra of different binaries with circumstellar
matter (e.g. Be-stars or algols) and their reconstructions from
disentangled spectra. The method consists in fixing the period of
orbit (1) equal to that of orbit (0) but converging either epochs
or periastron longitudes of both orbits together with the radial-velocity
amplitudes ($K$) as independent quantities. If the component 1 and 2
correspond to the primary and secondary star, the component 3 may also
be switched on, to correspond to (either emission or absorption) features
of the circumstellar matter. The amplitude and phase-shift of this component
are then correspond to the absolute value and orientation of the
superposition of the orbital and intrinsic velocity of the circumstellar
matter with respect to the center of mass of the system.
In principle, up to five component spectra corresponding to different
features corotating in the orbital plane of a binary may be treated
using KOREL, if the periods of all four orbits are fixed equal. Such
a disentangling always improved formally the fit, but usually it did
not provide a fully satisfactory explanation of a long-lasting series
of line-profiles. This may be explained by variations of the motion
and emissivity of the circumstellar matter on time-scales shorter then
the orbital period.

Sowers et al. (1998) used the method of Doppler mapping to
interpret the line-profiles of the H$_{\alpha}$ line of Cyg X-1.
They found a good agreement with observations if an emission source
attributed to a focused stellar wind is involved. Recently Jingzhi
Yan (2007) suggested to disentangle this focused stellar wind and
the primary component from the H$_{\alpha}$ line of Cyg X-1. The
preliminary results reported here are obtained by switching on
the components 1, 3 and 5 for the primary, the focused wind and
telluric water-vapor lines (which are quite strong here), resp.
The period of orbit (1) is set equal to that of orbit (0), orbit (3)
is the annual motion and orbit (2) is degenerated (cf. Fig.~\ref{obr05},
right).

The results are shown in Fig.~\ref{obr04} and Tab.~\ref{tab1}.
The profiles reconstructed from the disentangled components are
again superimposed on the observed 24 line-profiles plotted in
the chronological order from the top. The agreement of these
curves is surprisingly better than the one obtained for some
other interacting binaries with much less pronounced variability.
The agreement is a bit worse for the first two exposures taken
before the X-high episode, also compared to the last exposures,
where X-emission was low again and H$_{\alpha}$ emission high
(cf. Fig.~\ref{obr03}), but still with the P-Cyg profile.
The mean line-profile of the primary has a P-Cyg shape, the
component attributed to the focused stellar wind is a broad
emission. The disentangled spectrum of telluric lines is partly
contamined with the H$_{\alpha}$ emission, but again much less
than for many other emission-line binaries.
Both the primary and wind components are varying in strength,
but the analysis of this variability is postponed to a next
study based on more spectra disentangled with constrained
telluric component as described by Hadrava (2006).

\setlength{\unitlength}{1mm}
\begin{figure}[hbt]
\noindent\begin{picture}(127,65)
 \put(26,-13){\epsfxsize=75mm\epsfbox{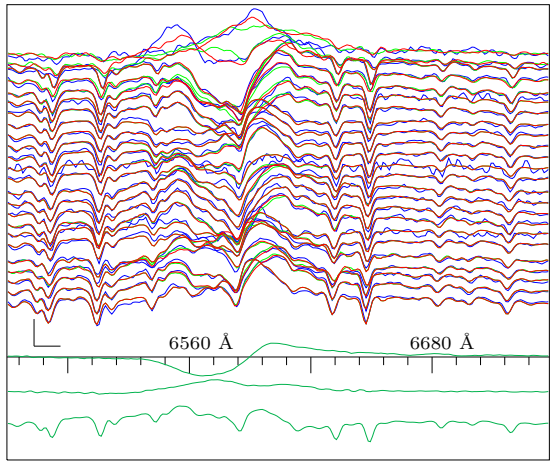}}
 \put(49.,14.7){\scriptsize 6560 \AA}
 \put(81.,14.7){\scriptsize 6680 \AA}
\end{picture}
\caption{
 \label{obr04} Disentangled H$_{\alpha}$-line}
\end{figure}

 The orbital parameters for the primary disentangled from
H$_{\alpha}$ are in good agreement with those obtained from
He-line here as well as in other studies. The radial-velocity
amplitude $K_{wind}=60.78\,{\rm km\,s}^{-1}$ and the phase shift
$\phi_{0}=0.46$ (with respect to the He-line) of the wind component
is not in a complete agreement with results given in text of Sowers
et al. (1998, p.~428)
who obtained $K_{wind}=68\,{\rm km\,s}^{-1}$ and $\phi_{0}=0.86$
using the tomographic method. Such a disagreement is not surprising,
because in spite of
some similarities, both methods are different, particularly in
taking into account the line-strength variability or the telluric
lines. Also the long-term stability of the focused-wind
component should be tested by additional spectra. However, the
present results indicate, that the velocity-distribution of
this component does not vary substantially before and during the
episode od high X-ray state.

\ack
This work has been done in the framework of the Center for
Theoretical Astrophysics (ref.~LC06014) with a support of grant
GA\v{C}R 202/06/0041. It is based on observational data obtained
using Ond\v{r}ejov 2-m telescope and ASM at RXTE satellite. The work
of teams at both these instruments is highly appreciated.

\pagebreak


\begin{thebibliography}{99}
\bibitem
 {}Gies~D.R., Bolton~C.T., Blake~M., Caballero-Nieves~S.M., Crenshaw~D.M.,
 Hadrava~P., Herrero~A., Hillwig~T.C., Howell~S.B., Huang~W., Kaper~L.,
 Koubsk\'{y}~P., McSwain~M.V., Melymuk~L. (2007), ``Stellar wind variations
 during the X-ray high and low states of Cygnus X-1", ApJ, submitted
\bibitem
 {}Hadrava~P. (1995), ``Orbital elements of multiple spectroscopic stars",
 A\&AS 114, 393
\bibitem
 {}Hadrava~P. (1997), ``Relative line photometry of eclipsing binaries",
 A\&AS 122, 581
\bibitem
 {}Hadrava~P. (2001), ``The method of spectra disentangling and its links
 to Doppler tomography" in: ``Astrotomography, indirect imaging methods in
 observational astronomy", eds. H.M.J. Boffin, D. Steeghs, J. Cuypers,
 Lecture notes in physics 573, 261 
\bibitem
 {}Hadrava~P. (2004), ``KOREL -- User's guide", Publ. Astron. Inst.
 ASCR 92, 15
\bibitem
 {}Hadrava~P. (2006), ``Disentangling of the spectra of binary stars --
 Principles, results and future development", Astrophys. \& Sp. Sc.
 304, 337
\bibitem
 {}Jingzhi Yan (2007), private communication
\bibitem
 {}Karas~V., Kraus~P. (1996), ``Doppler tomography of relativistic
 accretion disks", PASJ 48, 771
\bibitem
 {}Karitskaya~E.A., Shimanskii~V.V., Bochkarev~N.G., Sakhibullin~N.A.,
 Galazutdinov~G.A., Lee.~B.-C. (2007), ``Properties of the Cyg X-1 Optical
 Component", IAU Symp. 240, 130
\bibitem
{}Sowers~J.W., Gies~D.R., Bagnuolo~W.G., Jr., Shafter~A.W., Wiemker~R.,
 Wiggs~M.S. (1998), ``Tomographic analysis of H$\alpha$ profiles in
 HDE~226868/Cygnus X-1", ApJ 506, 424
\bibitem
{}\v{S}koda~P., \v{S}lechta~M. (2002), ``Reduction of spectra
 exposed by the 700mm CCD camera of the Ond\v{r}ejov telescope coud\'{e}
 spectrograph", Publ. Astron. Inst. ASCR 90, 22
\end{thebibliography}
\end{document}